
%
%
%



\font\openface=msbm10 at10pt
 %


\def\Reals{{\hbox{\openface R}}}

\def\implies{\Rightarrow}


 \def\dal{\displaystyle{{\hbox to 0pt{$\sqcup$\hss}}\sqcap}}


\def\lto{\mathop
        {\hbox{${\lower3.8pt\hbox{$<$}}\atop{\raise0.2pt\hbox{$\sim$}}$}}}
\def\gto{\mathop
        {\hbox{${\lower3.8pt\hbox{$>$}}\atop{\raise0.2pt\hbox{$\sim$}}$}}}
%
%
%




\def\isomorphic{\simeq}		


\def\to{\mathop\rightarrow}	

\def\tilde{\widetilde}		



\def\interior #1 {  \buildrel\circ\over  #1}     




\def\basisvector#1#2#3{
 \lower6pt\hbox{
  ${\buildrel{\displaystyle #1}\over{\scriptscriptstyle(#2)}}$}^#3}

\def\braces#1{ \{ #1 \} }






%
 \let\miguu=\footnote
 \def\footnote#1#2{{$\,$\parindent=9pt\baselineskip=13pt%
 \miguu{#1}{#2\vskip -7truept}}}
%
%

\def\BulletItem #1 {\item{$\bullet$}{#1}}


\def\AbstractBegins
{
 \singlespace                                        
 \bigskip\leftskip=1.5truecm\rightskip=1.5truecm     
 \centerline{\bf Abstract}
 \smallskip
 \noindent	
 } 
\def\AbstractEnds
{
 \bigskip\leftskip=0truecm\rightskip=0truecm       
 }

\def\ReferencesBegin
{
 \singlespace					   
 \vskip 0.5truein
 \centerline           {\bf References}
 \par\nobreak
 \medskip
 \noindent
 \parindent=2pt
 \parskip=6pt			
 }

\def\ref{\hangindent=1pc\hangafter=1} 

\def\section #1 {\bigskip\noindent{\bf #1 }\par\nobreak\smallskip\noindent}

\def\subsection #1 {\medskip\noindent[ {\it #1} ]\par\nobreak\smallskip}

\def\eprint#1{$\langle$#1\hbox{$\rangle$}}
 %

\def\linebreak{\hfil\break}


\def\author#1 {\medskip\centerline{\it #1}\smallskip}

\def\address#1{\centerline{\it #1}\smallskip}

\def\furtheraddress#1{\centerline{\it and}\smallskip\centerline{\it #1}\smallskip}



\def\printVersionNumber{\rightline{version 4.1}} 
\def\printVersionNumber{}



 \message{Assuming 8.5 x 11 inch paper.}
 \message{' ' ' ' ' ' ' '}

\magnification=\magstep1	          
\raggedbottom

%

\voffset=0.6 true in

\parskip=9pt

\def\singlespace{\baselineskip=12pt}      
\def\sesquispace{\baselineskip=16pt}      



\def\Omegatilde{{\tilde\Omega}}
\def\mutilde{{\tilde\mu}}
\def\Rtilde{{\tilde{\cal R}}}
\def\Ctilde{{\tilde{C}}}
\def\Stilde{{\tilde{S}}}
\def\R{{\cal R}}
\def\Rstem{\R_S}	
\def\follows{\succ}		


\phantom{}
\vskip -1 true in
\medskip

\printVersionNumber


\rightline{gr-qc/0202097}
\rightline{SU--GP--01/12--1}	   

\vskip 0.3 true in

\bigskip
\bigskip

\sesquispace
\centerline{ {\bf General Covariance and 
   the ``Problem of Time'' in a Discrete Cosmology}\footnote{*}%
{ To appear in the proceedings of the Alternative Natural Philosophy
  Association meeting, held August 16-21, 2001, Cambridge, England.
  This is the slightly extended text of a talk presented at the meeting
  by R.D.S. and based on joint work of the authors. }}

\bigskip


\singlespace			        

\author{Graham Brightwell}
\address{Mathematics Department, London School of Economics}

\author{H.~Fay Dowker}
\address{Physics Department, Queen Mary College, University of London}

\author{Raquel S. Garc{\'\i}a}
\address{Physics Department, Imperial College, London}

\author {Joe Henson}
\address{Physics Department, Queen Mary College, University of London}

\author{Rafael D. Sorkin}
\address{Physics Department, Queen Mary College, University of London}
\furtheraddress{Department of Physics, Syracuse University}

\AbstractBegins 
Identifying an appropriate set of ``observables'' is a nontrivial task
for most approaches to quantum gravity.  We describe how it may be
accomplished in the context of a recently proposed family of stochastic
(but classical) dynamical laws for causal sets.  The underlying idea
should work equally well in the quantum case.
\AbstractEnds


\sesquispace

\section {1. Introduction}
Perhaps I should begin by clarifying what I {\it don't} mean by the
quoted phrase ``problem of time'' in my title.  In the canonical
quantization of gravity, as it is normally understood (and sought) the
fundamental object of attention is not spacetime but space alone
(corresponding to something like a Cauchy surface in spacetime) and
anyone following this approach is sooner or later faced with the problem
of {\it recovering time} from a frozen formalism (as it's sometimes
called) from which time as such is absent.  The problem of time {\it in
this sense} is, I believe, insoluble and it will not be the subject of
this talk.  (See [1] [2].)
Indeed, since I will be presupposing that the deep structure of
spacetime is that of a {\it causal set}, temporality will be built in at
the most fundamental level, and there will be no need to ``recover''
it, just as there would be no need to ``recover time'' in a continuum path
integral approach to quantum gravity based on  the Lorentzian manifold as
the fundamental structure.\footnote{$^\star$}
{Not that, in saying this, I mean to downplay the need to explain why
 the dynamics tends to favor that small minority of causal sets which
 resemble Lorentzian manifolds over the multitude of those which don't.}

Nevertheless, there still remain vital interpretational issues related
to the generally covariant nature of gravity that sometimes are also
called ``problems of time'',\footnote{$^\dagger$}
{Observe in this connection that it is the diffeomorphism invariance of
 the gravitational Lagrangian (or rather action-integral) which is
 responsible for the ``frozen'' character of the corresponding canonical
 formalism}
issues that one can point to by asking (in the
language of one version of quantum mechanics) ``What are the observables
of quantum gravity?''.  For my part, I'd prefer a reference to
``be-ables'' rather than ``observe-ables'', but the question I will be
addressing in this talk is best posed without using either
word.  Rather we can simply ask ``To what questions (about the causal
set) can the dynamics give answers?''

Within the (stochastic but still classical) dynamical framework we will
be considering, a {\it question} (in logicians' language a
``predicate'') corresponds to a collection of causal sets (the
predicate's ``extension'') and its answer is not a simple yes or no, but
the {\it probability} that the answer will be yes, which mathematically
is the {\it measure} of the corresponding collection. Thus, we will be
asking which classes of causal sets (the ``histories'' of the theory)
are {\it measurable} in a way compatible with general covariance.  I
will propose a very definite answer to this question that will follow
naturally from the way the dynamics will be defined as a stochastic
process.  Not all the resulting measurable classes (or their associated
predicates) will obviously possess an accessible physical meaning, but a
certain subset will (those based on ``stems'').  Fortunately this subset
is very big, and it seems quite possible that nothing of interest --- or
even nothing at all --- remains outside of it.  After describing a
precise conjecture to this effect, I will conclude by asking whether a
proof of the conjecture would settle all the interpretational issues, or
whether a logically independent notion of ``conditional probability'' is
also needed.  Before I can present these thoughts more fully, however,
some review is needed of the causal set idea itself.

\section {2. A Brief review of causal set theory}
The causal set hypothesis states that the deep structure of spacetime is
that of a discrete partial order, and that, consequently, ``quantum
gravity'' can be realized only as a quantum theory of causal sets
(``causets'' for short).  To say what a causet is structurally is easy:
it's a locally finite partial order; but to specify fully the meaning of
the words ``quantum theory of causets'' is much harder.  
It seems plausible that the dynamics of such a theory (its ``laws of
motion'' if you will) would be specified mathematically in terms of a
``decoherence functional'' [3] or ``quantum measure''
[4] but we are only beginning to understand the principles
that might lead us to the correct one.\footnote{$^\flat$}
{As I am using it, the term ``decoherence functional'' denotes no more
 than a certain type of mathematical object, formally defined by axioms
 of bi-additivity, etc.  In particular, I do not mean to add any
 requirement that it actually ``decohere'' in the sense of being
 diagonal on any partition of the ``sample space'' $\Omega$ to be
 defined below.}
On the other hand, we do have a family of
{\it classical} dynamical laws derived from well defined general
principles including a principle of ``discrete general covariance'' and
a certain principle of ``Bell causality''.  To the extent that these
principles carry over to the quantal case, we are thus a considerable
way along the road to a quantum dynamics for causets.  Moreover, the
projected quantum dynamics shares enough attributes with the existing
classical one that it seems worthwhile to consult the latter for
indications of what we can expect from its quantum generalization.

In this way, it has been possible to make some guesses and heuristic
predictions which have begun to bring the theory into contact with
phenomenology.  The list of these must include first of all the
anticipation [5] [2] [6] 
of time-dependent fluctuations in the cosmological
constant whose predicted current magnitude of $10^{-120}$ in
natural units has turned out to be in good 
accord with recent observations.  Second
[7] there exists a (purely kinematical) counting of
``horizon quanta'' whose number is (both in order of magnitude and
proportionality to horizon area) compatible with the Bekenstein-Hawking
formula for both equilibrium {\it and} nonequilibrium examples of black
holes.  We also have an indication [8] of how some of the
notorious large numbers of cosmology might be explained, as well as a
framework [9]
within which Hawking radiation can be addressed.  (In the way
of practical tools there also exists an extensive library 
[10] of Lisp functions designed for working with causets,
i.e. basically with finite ordered sets or ``posets''.)

In the following, however, I wish to concentrate not on the
phenomenological aspects of the theory but on interpretational ones,
more specifically on the question already raised of which physically
meaningful (i.e. generally covariant) predicates correspond to classes
of causets to which the dynamics can assign a measure (probability).  To
this end, I will have to describe more precisely the family of
stochastic dynamical laws on which these considerations will be based.

\section {3. The classical (stochastic) dynamics of sequential growth and its formal
definition as a stochastic process}
As indicated by the title of this talk, we will work within a
``cosmological'' setting, in the sense that the probabilities in
question will pertain to the causet as a whole, not just to some part of
it.  This seems to be necessary, as it is difficult to imagine how any
generally covariant procedure could single out a definite ``subregion''
of the universe in an {\it a priori} manner.  A cosmological standpoint
is also appropriate formally, since the dynamical laws we will be using
conceive of the ``time-development'' of the causet as a process of
sequential growth in which elements appear (``are born'') one by one;
and, as formulated, these laws make sense only if there is a genuine
``beginning condition'' in which there are no elements at all (or at
most a finite number).

The family of dynamical laws in question is described in detail in
[11], where it is derived as the unique (generic) solution of
certain conditions of ``internal temporality'', ``Bell causality'' and
``discrete general covariance''.  For present purposes, it is enough to
know that the resulting scheme describes a stochastic birth process
which, ``at stage $n$'', yields a poset $\Ctilde_n$ of $n$ elements,
within which the most recently born element is maximal.  If one employs
a genealogical language in which ``$x\prec{y}$'' can be read as ``$x$ is
an ancestor of $y$'', then the $n^{th}$ element (counting from $0$) must
at birth ``choose'' its ancestors from the elements of $\tilde{C}_n$,
and for consistency it must choose a subset $S$ with the property that
$x\prec{y}\in{S}{\,\implies\,}x{\in}S$.  (Every ancestor of one of my
ancestors is also my ancestor.)  Such a subset $S$ (which is necessarily
finite) will be called a {\it stem}.  The dynamics is then determined
fully by giving the {\it transition probabilities} governing each such
choice of 
$S\subseteq\Ctilde_n$.
(To understand how a set of probabilities can
determine a dynamics, think of a random walk.  The ``law of motion'' of
the walker is specified by giving, for each possible time and location,
the probability of taking, say, a step to the right, a step to the left,
or just staying put.)

We can formalize this scheme by introducing for each integer
$n=0,1,2,\cdots$ the set $\Omegatilde(n)$ of {\it labeled} causets of
$n$ elements.  By definition, a member of $\Omegatilde(n)$ is thus a
set $\Ctilde_n$ with cardinality $n$ carrying a relation $\prec$ such that 
$x{\prec}y{\prec}z{\,\implies\,}x{\prec}z$ (transitivity), and 
$x{\not\prec}x$ (irreflexivity), and whose elements are labeled by
integers $0,1,\cdots,n-1$ that record their order of birth.  Moreover
this labeling is {\it natural} in the sense that
$x{\prec}y{\,\implies\,}l(x)<l(y)$, $l(x)$ being the label of $x$.  Each
birth of a new element 
occasions one of the allowed transitions from $\Omegatilde(n)$ to
$\Omegatilde(n+1)$ and occurs with a specified conditional probability
$\tau$
(which turns out to depend only on a pair of simple invariants of the
ancestor set $S\subseteq\Ctilde_n$ 
of the newborn).  


Mathematically, however, such a set of transition probabilities $\tau$
does not yet qualify as a stochastic process, and for good reason.  In
the presence of fundamental randomness, no certain predictions are
possible.  Instead, the ``laws of motion'' can at best give
probabilistic answers to questions about what the object under study
(causet, random walker, etc.) will do, answers of the form ``Yes, with
probability $p$'', where $p\in[0,1]$.  For example, in the case of a
random walk on the integers, one might ask ``Will the walker ever return
to the origin?''.  Since the return, if it occurs at all, might be
postponed to an arbitrarily late time, its probability $p$ can be given
meaning only in terms of a limiting process.  For this particular
question, however, there exists an obvious definition of $p$ as the
limit of an increasing sequence of probabilities, each of which can be
computed from a finite number of elementary transition probabilities
$\tau$.  On the other hand, for a question like ``Will the sequence of
locations of the walker form an irrational decimal fraction when reduced
modulo ten?'', it is not immediately clear whether any meaningful
probabilistic answer can be given at all on the basis of the $\tau$.

In order to arrive at a definite theory, then, one needs to specify 
the set of questions that the theory can answer and for each one
of them, explain how in principle, the  ``yes'' probability can
be computed.  Fortunately, there exists a standard construction which
will accomplish both these tasks starting from any consistent set of
transition probabilities $\tau$.  The result of this construction is a
triad consisting of a 
{\it sample space} $\Omega$, 
a {\it $\sigma$-algebra} $\R$ on $\Omega$,
and a {\it probability measure} $\mu$ with domain $\R$.
In relation to the above two tasks, each member $Q$ of $\R$ corresponds to
one of the answerable questions and its measure $p=\mu(Q)$ is the
answer.  Such a triad constitutes what one means mathematically by a
``stochastic process'',  the transition probabilities $\tau$ serve
only as raw material for its construction.
(That $\R$ is a $\sigma$-algebra on $\Omega$ means that it is
 a family of subsets of $\Omega$ closed under complementation and
 countable intersection.  
 That $\mu$ is a probability measure with domain $\R$ means that 
 it takes members of $\R$ to non-negative real numbers and is 
 $\sigma$-additive, with $\mu(\Omega)=1$.  
 Finally, $\sigma$-additivity means that $\mu$ assigns to the union
 of a countable collection of mutually disjoint sets in its domain the
 sum of the measures it assigns to the individual sets.)

In the case at hand, the sample space is the set
$\Omegatilde=\Omegatilde(\infty)$ of {\it completed labeled causets}
these being the infinite causets that would result if the birth process
were made to ``run to completion''.  (I'll use a tilde to indicate
labeling.  Notice that a completed causet, though infinite, is still
locally finite.\footnote{$^\star$}
{Local finiteness, a formal realization of the concept of discreteness,
 is the property that the {\it order interval},
 $\braces{x|{a}\prec{x}\prec{b}}$, is finite for all elements $a$ and
 $b$.}
Indeed it is past-finite in the sense that no element has more than a
finite number of ancestors.)  The dynamics is then given by a
probability measure $\mutilde$ constructed from the $\tau$ whose domain
$\Rtilde$ is a $\sigma$-algebra which I will specify more fully in a
moment. 
For future use, we will need 
in addition to $\Omegatilde$
the corresponding space $\Omega$ of
completed {\it unlabeled} causets, whose members can also be viewed in
an obvious manner as equivalence classes within $\Omegatilde$.

(To be pedantically precise, one should perhaps speak of the members of
 $\Omega$ and $\Omegatilde$ not as single causets but as isomorphism
 equivalence classes of them --- what one might call ``abstract
 causets''.)

At first hearing, calling a probability measure a dynamical law might
sound strange, but in fact, once we have the measure $\mutilde$ we can
say everything of a predictive nature that it is possible to say {\it a
priori} about the behavior of the causet $C$.  For example, one might ask
``Will the universe recollapse?'' (a question analogous to our earlier
question of whether the random walker would return to the origin).
Mathematically, this is asking whether $C$ will develop a ``post'',
defined as an element whose ancestors and descendants taken together
exhaust the remainder of $C$.  Let $A\subseteq\Omegatilde$ be the set of
all completed labeled causets having posts.\footnote{$^\dagger$}
{One can show that $A\in\Rtilde$, so that $\mutilde(A)$ is defined.}
Then our question is equivalent to asking whether $C\in{A}$, and the
answer is ``yes with probability $\mutilde(A)$.''    
It is thus $\mutilde$ that expresses the ``laws of motion'' (or better
``laws of growth'') that constitute our stochastic dynamics: its domain
$\Rtilde$ tells us which questions the laws can answer, and its values
$\mutilde(A)$ tell us what the answers are.

Of course, this sketch of how a sequential growth model is built up is
incomplete, because I haven't explained the construction that leads from
the transition probabilities $\tau$ to the measure $\mutilde$.  The full
details of this construction can be found in many textbooks of
probability theory (e.g. [12]), but for present purposes, all we
really need to know is how the domain $\Rtilde$ of $\mutilde$ is
obtained.  
To each finite causet $\Stilde\in\Omegatilde(n)$ one can associate the
so called ``cylinder set'' comprising all those $\Ctilde\in\Omegatilde$
whose first $n$ elements (those labeled $0\cdots{n-1}$) form an
isomorphic copy of $\Stilde$; and $\Rtilde$ is then the smallest
$\sigma$-algebra containing all these cylinder sets.
More constructively, $\Rtilde$ is the collection of all subsets of
$\Omegatilde$ which can be built up from the cylinder sets by a
countable process involving union, intersection and
complementation.\footnote{$^\flat$} 
{A slightly bigger $\sigma$-algebra than $\Rtilde$ can be obtained by
 adjoining the sets of $\mutilde$-measure zero, but what these sets are
 will depend in general on the specific dynamical law, as determined,
 e.g., by a choice of the parameters $t_n$ of [11].}

Finally, what about the transition probabilities $\tau$ themselves, on
which the whole construction is based?  Modulo certain non-generic
solutions, the possibilities for the $\tau$ have been classified in
[11], the main conclusion being that $\tau$ generically takes the
form $\tau=\lambda(\varpi,m)/\lambda(n,0)$ where, for the potential
transition in question, $\varpi$ is the number of ancestors of the new
element, $m$ the number of its ``parents'', and $n$ the number of
elements present before the birth, and where $\lambda(\varpi,m)$ is
given by the formula $\sum_k {\varpi-m\choose k-m}t_k$ with the $t_k$
being the free parameters or ``coupling constants'' of the theory.  (For
more details see [11] or [13].)

\section {4. Two meanings of general covariance}
It might seem strange that our growth law has been expressed in terms
of {\it labeled} causets.  After all, labels in this discrete setting
are the analogs of coordinates in the continuum, and the first lesson of
general relativity is precisely that such arbitrary identifiers must be
regarded as physically meaningless: the elements of spacetime --- or of
the causet --- have individuality only to the extent that they acquire
it from the pattern of their relations to the other elements.  It is
therefore natural to introduce a principle of ``discrete general
covariance'' according to which ``the labels are physically
meaningless''.

But why have labels at all then?  For causets, the reason is that we
don't know otherwise how to formulate the idea of sequential growth, or
the condition thereon of Bell causality, which plays a crucial role in
deriving the dynamics [11].  Ideally perhaps, one would formulate
the theory so that labels never entered, but so far, no one knows how to
do this --- anymore than one knows how to formulate general relativity
without introducing extra gauge degrees of freedom that then have to be
canceled against the diffeomorphism invariance. 

Given the dynamics as we {\it can} formulate it, discrete general
covariance plays a double role.  On one hand it serves to limit the
possible choices of the {\it transition probabilities} in such a way
that the labels drop out of certain ``net probabilities'', a condition
made precise in [11].  This is meant to be the analog of requiring
the gravitational action-integral $S$ to be invariant under
diffeomorphisms (whence, in virtue of the further assumption of
locality, it must be the integral of a local scalar concomitant of the
metric).  On the other hand, general covariance limits the {\it
questions} one can meaningfully ask about the causet (cf. Einstein's
``hole argument'' [14]).  
It is this second limitation
that is related to the ``problem of time'', and it is only this aspect
of discrete general covariance that I am addressing in the present talk.

Just as in the continuum the demand of diffeomorphism-invariance makes
it harder to formulate meaningful statements,\footnote{$^\star$}
{Think, for example, of the statement that light slows down when passing
 near the sun.}
so also for causets the demand of discrete general covariance has the
same consequence, bringing with it the risk that, even if we succeed in
characterizing the covariant questions in abstract formal terms, we may
never know what they mean in a physically useful way.  I believe that a
similar issue will arise in every approach to
quantum gravity, discrete or continuous (unless of course general
covariance is renounced).\footnote{$^\dagger$}
{In the case of canonical quantum gravity, this issue {\it is} the
 problem of time.  There, covariance means commuting with the
 constraints, and the problem is how to interpret quantities which do so
 in any recognizable spacetime language.  For an attempt in string
 theory to grapple with similar issues see [15].}
However, it seems fair to say that both the nature
of the difficulty and the manner of its proposed resolution 
will
appear with
special clarity in the context of causal sets, whose discreteness
removes many of the technical difficulties that tend to obscure the
underlying physical issues in the continuum.

\section {5. What are the covariant questions?}
Given the formal developments of Section 4, it is not hard to see which
members of $\Rtilde$ express covariant predicates, and from a dynamical
point of view, these are the only covariant predicates of interest.

Before describing them, let me illustrate
the issue we face with the question, ``Which (unlabeled) causet is
formed by the first $n$ elements to be born?''.  In effect, we are
asking for the probability distribution induced by our measure
$\mutilde$ on the space $\Omega(n)$ of unlabeled $n$-orders, but,
although such a distribution can be computed, it has no obvious meaning
because a question like ``Do the first three elements of $C$ form a
3-chain\footnote{$^\flat$}%
{A (finite) chain is a causet whose elements can be arranged so that
 each is an ancestor of the next.  For a 3-chain, we have three elements
 $a$, $b$, and $c$ such that $a\prec{b}\prec{c}$.}?''
has in general different answers depending on what order of birth you
impute to the elements of $C$.  Thus if we were to divide the set
$\Omegatilde$ into two parts, the ``yes'' part composed of those
$\Ctilde\in\Omegatilde$ whose first 3 elements make up a chain, and the
``no'' part composed of those $\Ctilde$ whose first 3 elements make up
one of the other four 3-orders, then some members of the ``yes'' set
would be isomorphic to members of the ``no'' set. 

We see now what it means for a subset of $A\subseteq\Omegatilde$ to be
covariant: it cannot contain any labeled completed causet $\Ctilde$
without containing at the same time all those $\Ctilde'$ isomorphic to
$\Ctilde$ (i.e. differing only in their labelings).  To be measurable as
well as covariant, $A$ must also belong to $\Rtilde$.  
Let $\R$ be the collection of all such sets:
$A\in\R{\iff} {A}\in\Rtilde
\ {\rm and}\ 
\forall \Ctilde_1 \isomorphic \Ctilde_2\in\Omegatilde,
\Ctilde_1\in{A}\implies\Ctilde_2\in{A}$.
It is not hard to see that $\R$ is a sub-$\sigma$-algebra of $\Rtilde$,
whence the restriction of $\mutilde$ to $\R$ is a measure $\mu$ on the
space $\Omega$ of unlabeled completed causets.\footnote{$^\star$}
{As just defined, an element $A\in\R$ is a subset of $\Omegatilde$.
 However, because it is re-labeling invariant, it can also be regarded
 as a subset of $\Omega$, an equivalence which I will henceforth utilize
 without explicit mention.}
It is this measure $\mu$ that provides the answers to all the covariant
questions for which the dynamics has answers.\footnote{$^\dagger$}
{Notice the distinction that arises here between a subset of
 $\Omegatilde$ that fails to belong to $\R$ and one that fails even to be
 covariant (one that cannot be regarded as a subset of $\Omega$).
 The former corresponds to a question that the dynamics can't answer,
 the latter to a question that is without any physical meaning
 at all.}
But what do these questions signify physically?

\section {6. Stem sets and a conjecture}
Among the questions belonging to $\R$ there are some which do have a
clear significance.  
Let $S\in\Omega(n)$ be any finite unlabeled causet and let
$R(S)\subseteq\Omega$ be the ``stem set'',
$\braces{C\in\Omega|C \,\hbox{admits}\, S \,\hbox {as a stem} }$.
(Thus $R(S)$ comprises those unlabeled completed causets with the
property that, with respect to some natural labeling, the first $n$
elements form a causet isomorphic to $S$.)  Since (as one can prove)
$R(S)$ is measurable, it belongs to $\R$.  
For this particular element of $\R$, the meaning of the corresponding
causet question is evident: 
``Does the causet possess $S$ as a stem?''.\footnote{$^\flat$} 
{Here is a simple example.  Let $\Lambda$ be the 3-element causet given
 by $a\prec{c},\, b\prec{c}$, let $V$ be its dual (given by
 $a\follows{c},\, b\follows{c}$), and let $S$ be the 2-chain (given by
 $a\prec{b}$).  Then $V$ possesses $S$ as a stem, while $\Lambda$ does
 not.}
Equally evident is the significance of any question built up as a
logical combination of stem-questions of this sort.  To such
compound stem-questions belong members of $\R$ built up from stem-sets
$R(S)$ using union, intersection and complementation (corresponding to
the logical operators `or', `and' and `not').  If all the members
of $\R$ were of this type, we would not only have succeeded in
characterizing the dynamically meaningful covariant questions at a
formal level, but we would have understood their physical significance
as well.\footnote{$^\star$}
{This is not yet their {\it phenomenological} meaning, of course.  To
 get the phenomenological significance (in cases where there is one) one
 must translate between the combinatorial language proper to the causet
 and the geometrical and field-theoretic language of macroscopic
 physics.}
The following conjecture asserts that, to all intents and
purposes, this is the case.

\noindent
{\bf Conjecture}
The ``stem-sets'' $R(S)$ generate\footnote{$^\dagger$}
{A family ${\cal F}$ of subsets is said to {\it generate} a
 $\sigma$-algebra ${\cal A}$ if ${\cal A}$ is the smallest
 $\sigma$-algebra containing all the members of ${\cal F}$.  
 For example, the cylinder sets introduced above generate the
 $\sigma$-algebra $\Rtilde$.}
the $\sigma$-algebra $\R$ up to sets of measure zero.

\noindent
Here the 
technical qualification about sets of measure zero complicates the
statement of the conjecture but does not essentially weaken it.
Some such qualification is required, unfortunately, because one can
exhibit counterexamples to the unqualified conjecture.
Notice, once again, that the words ``measure zero'' have meaning only
with respect to some choice of 
stochastic dynamical law, 
since different
choices correspond to different measures $\mu$ which, in general will
possess different families of measure-zero sets.  Thus, a more complete
phrasing of the conjecture would read ``For every choice of sequential
growth model, the stem-sets $R(S)$ generate $\R$ up to sets of measure
zero.''  Here a sequential growth model is any member of the ``generic''
family described briefly in Section 3 above (and at length in
[11]), and more generally any solution of the conditions of Bell
causality, etc. delineated in [11].

The conjecture asserts that $\Rstem=\R$, where $\Rstem$ is the
subalgebra of $\R$ generated by the stem-sets $R(S)$.  In a
moment, I'll present some evidence in favor of this equality, but first
I'd like to stress that, even if $\Rstem$ fails to exhaust $\R$, it
still supplies us with a large store of predicates whose physical
significance is transparent, and this store probably suffices for
practical purposes since one's experience so far indicates that all
predicates of interest belong to it, either outright or up to a set of
measure zero. 
For example, the predicate ``contains a post'' belongs to $\Rstem$.

Now the intuitive import of our conjecture is that everything we need to
say about a causet can be phrased in terms of its stems.  If this is so,
then we'd expect at least that a specification of the full set of
stems of a given causet $C\in\Omega$ would suffice to characterize $C$
uniquely.  As a matter of fact, not all causets enjoy this feature, but
the exceptions are rare enough that the probability of one of them being
produced by a sequential growth process is zero.  (Employing the
jargon of probability theory, we may say that a causet $C$ produced by
one of the sequential growth processes is {\it almost surely}
characterized by its stems.)  That this is so speaks in favor of the
conjecture but, as far a I know, is not enough to demonstrate it fully.
Its proof, in any case, can be given in three steps as follows
[16].

\item{(a)} If $C\in\Omega$ {\it fails} to be characterized by its stems
then it must contain an infinite number of copies of some stem $S$.

\item{(b)} If $C\in\Omega$ contains an infinite number of copies of the
same stem $S$ then it must contain an infinite antichain (indeed an
infinite level).

\item{(c)} The probability is zero that a $C$ produced by one of the
classical sequential growth models will contain an infinite level, or
indeed any infinite antichain.

\noindent
Strictly speaking, statement (c) has been proven only for the models
described above which are parameterized by the ``coupling constants''
$t_0, t_1, t_2,\cdots$, but these are generic in the sense that they
include all solutions for which no transition probability $\tau$
vanishes [11].
Moreover (c) fails for the special case where $t_n=0$ for all $n\ge{2}$.
But this exception does not affect the main conclusion, because if $t_0$
is the only nonzero $t_n$ then $C$ is almost surely an infinite
antichain, while if $t_0$ and $t_1$ are both nonzero then $C$ is almost
surely an infinite number of copies of the tree in which each element
has an infinite number of children; and both these causets are also
characterized by their stems.

To get a feel for what these results mean, consider the causet $C'$
consisting of an infinite number of unrelated copies of the infinite
chain
 $e_0 \prec e_1 \prec e_2 \prec \cdots $ .
This causet is {\it not} characterized by its stems, because, e.g., the
causet $C''$ made from $C'$ by adjoining a single $n$-chain
($1\le{n}<\infty$) has precisely the same stems as $C'$.  (A causet $S$
is such a stem iff it is the sum of a finite number of finite chains.)
The proof outlined above then tells us that $C'$ and $C''$ are
infinitely unlikely to ``grow'' in any of the models with $t_2>0$.

\section {7. Conclusion and some further questions}
General covariance in classical gravity is a two-edged sword.  
On one hand it is both philosophically satisfying (at least to those of
us who favor ``relational'' or ``dialectical'' theories) and
heuristically fruitful for the way in which it limits the possible
equations of motion.
On the other hand it forces the consequent formalism farther from
experience, because it renders meaningless all statements which are not,
at least implicitly, of a global character.  (In this sense, the context
for a generally covariant theory is always cosmological in scope.)  The
meaningful statements become thereby both harder to formulate and harder
to interpret (``problem of time'').

In the case of (classical) general relativity, these difficulties are
somewhat mitigated by the theory's determinism, which means that
predictions can effectively be made and verified locally.  Such is not
the case, however, in an indeterministic theory, where the dynamics must
presumably be expressed via relations of a probabilistic character among
covariant statements whose global character is much harder to
circumvent.  The problem of characterizing and interpreting such
statements must therefore arise in virtually every approach to quantum
gravity, including of course the causet approach. \footnote{$^\flat$}
{Among the possible exceptions would be approaches which attempt to
 restore strict determinism.}

But the classical sequential growth models of causet dynamics are also
indeterministic, and consequently the same problems arise for them.
These models, which are intended primarily as a (practical and
conceptual) stepping stone to the quantum theory of causets, offer a
precisely defined schema within which the interpretational difficulties
deriving from general covariance can be confronted.  And within this
setting, we have largely resolved them: the dynamically meaningful
predicates have been precisely characterized in Section 5, and an
interpretation for many or all of them (in terms of ``stem-predicates'')
has been set forth in Section 6.

For an ordinary stochastic process unfolding against the background of
some non-dynamical parameter time, the predicates corresponding to
measurable sets of trajectories can all be built up as logical
combinations of more elementary ones that acquire their full meaning in
a finite time. (They don't refer to the infinite future).  So it is also
for the ``cylinder set'' predicates introduced in Section 3, but these
predicates are unfortunately meaningless {\it per se} since they are not
generally covariant. 
The nearest covariant replacements for these cylinder sets
would seem to be the stem-sets $R(S)$.  Unlike the cylinder sets,
they do refer to the infinite future, for, although the predicate
corresponding  to a stem-set can become true in a finite time (after the
birth of a finite number $n$ of elements) it can only strictly speaking
become false in the limit $n\to\infty$.\footnote{$^\star$}
{Even if a stem $S$ has not appeared yet, we can never be absolutely
 certain that it won't appear later on.}
Nonetheless, falsehood can become certain ``for all practical purposes'', 
and in this sense, the truth or falsehood of a stem-predicate is also
``verifiable in a finite time''.

Moreover the physical significance of a stem-predicate is transparent,
since its truth is decided simply by whether the ``universe'' (i.e. the
actual causet) does or does not (never will) contain the stem in
question.  To the extent, then, that all assertions of dynamical
interest can be built up from the stem predicates, we will have resolved
the ``problem of time'' for this range of models.  A proof of the
conjecture of Section 6 would vouchsafe us this conclusion.  It is
clear, moreover, that this relatively satisfactory situation depends
heavily on the discreteness of the causal set.

To what extent can we hope to repeat these same steps in the quantum
case (the case of quantum gravity)?  As a mathematical object, a quantum
measure/decoherence functional is not that different from a classical
measure like $\mu$ above, so one might hope much of the above would go
through relatively unchanged.  Two hurdles arise immediately, however.
First, of course, the predictive meaning of a quantum measure is much
less well understood than that of a classical probability measure,
whence, even if we could say precisely which subsets of $\Omega$ were
``measurable'', we still would not know exactly how to use this
information.\footnote{$^\dagger$}
{For an attempt to codify the predictive use of the quantum measure
 without first reducing it to a set of classical probabilities, see
 [17].}
Second, the theorems that above led us from the transition probabilities
$\tau$ to the measure $\mutilde$ (and thence to $\mu$) break down in the
quantum case because the complex amplitudes that one is constructing are
no longer necessarily bounded in absolute value, unlike probabilities
which are confined to the compact space $[0,1]\subseteq\Reals$.  How
serious these problems are is hard to say, but certainly the second of
them means that further technical developments would need to occur
before one could rigorously state a quantum analog of the conjecture of
Section 6.

Finally, I want to raise --- without attempting to settle it --- a
question about the status of conditional probability in quantum gravity,
or rather in the classical analog of causet quantum gravity which has
been the basis of our considerations here.  Certainly, one can
``relativize'' the measure $\mu$ to any measurable subset of $\Omega$
and thereby lend meaning to questions of the sort ``Given that $S_1$
occurs as a stem in $C$ (the universe), what is the probability that
$S_2$ also occurs as a stem?''.  If such questions are the only ones we need to
consider, then conditional probabilities here will have the same derived
status as they do in other branches of probability theory.  But how clear
is it that such a syntax captures what we'd really like to ask?
Consider instead something like this: ``Given that $S_1$ occurs as a
stem in $C$ what is the probability that a second stem $S_2$ occurs
containing {\it this particular copy of} $S_1$ (the $S_1$ that ``we
inhabit'')?''.   When $C$ contains more than one stem isomorphic to
$S_1$, this second type of question seems different from the first, and
indeed not even clearly defined on the sole basis of the ``absolute''
measure $\mu$.  Does this mean we need a {\it logically independent}
concept of conditional probability and an extended formalism to express
it, which would re-open the ``problem of time'' in a new context?  Or is
it enough to remark that stems of sufficient complexity are unlikely to
occur more than once, whence the problem is absent in practice?  As
with other conceptual issues raised by quantum gravity, so also with
this ``this particular stem'' issue, it's hard to say whether its
resolution demands deep thought or just a bit of progress on the
technical front.


\bigskip\noindent

This research was partly supported by NSF grant PHY-0098488, by a grant
from the Office of Research and Computing of Syracuse University, and by
an EPSRC Senior Fellowship at Queen Mary College.  I would also like to
express my gratitude to Goodenough College, which provided an estimable
environment for life and work during my stay in London, where this paper
was written.

\ReferencesBegin

\ref
[1] 
C.J.~Isham, ``Canonical Quantum Gravity and the Problem of Time'',
  in L.~A. Ibort and M.~A.~Rodriguez (eds.), 
  {\it Integrable Systems, Quantum Groups, and Quantum Field Theories}
  (Kluwer Academic Publishers, London, 1993)
   pp. 157--288
  \eprint{gr-qc/9210011} ;
\linebreak
K.V.~Kucha{\v{r}}, ``Canonical Quantum Gravity''  
    in R.J.~Gleiser, C.N.~Kozameh, O.M. Moreschi (eds.),
    {\it ``General Relativity and Gravitation 1992 '': Proceedings of the
     Thirteenth Conference on General Relativity and Gravitation}, 
     held Huerta Grande, Cordoba, 28 June-4 July, 1992
     (Bristol, IOP Publishing 1993) .
    %

\ref
[2]
R.D.~Sorkin,
``Forks in the Road, on the Way to Quantum Gravity'', talk 
   given at the conference entitled ``Directions in General Relativity'',
   held at College Park, Maryland, May, 1993,
   {\it Int. J. Th. Phys.} {\bf 36}: 2759--2781 (1997)   
   \eprint{gr-qc/9706002}
   %

\ref
[3]  
J.B.~Hartle, ``Spacetime Quantum Mechanics and the Quantum 
 Mechanics of Spacetime'',
 in B.~Julia and J.~Zinn-Justin (eds.),
 {\it Les Houches, session LVII, 1992, Gravitation and Quantizations}
 (Elsevier Science B.V. 1995)
 \eprint{gr-qc/9304006}.


\ref
[4]  
R.D.~Sorkin,
``Quantum Mechanics as Quantum Measure Theory'',
   {Mod. Phys. Lett. A} {\bf 9}:3119-3127 (No.~33) (1994)
   \eprint{gr-qc/9401003}. 
\linebreak
Roberto B.~Salgado, ``Some Identities for the Quantum Measure and its 
Generalizations'',
 \eprint{gr-qc/9903015}.

\ref
[5] 
R.D.~Sorkin, 
``Spacetime and Causal Sets'', 
     in J.C. D'Olivo, E. Nahmad-Achar, M. Rosenbaum, M.P. Ryan, 
              L.F. Urrutia and F. Zertuche (eds.), 
    {\it Relativity and Gravitation:  Classical and Quantum} 
    (Proceedings of the {\it SILARG VII Conference}, 
      held Cocoyoc, Mexico, December, 1990), 
    pages 150-173
    (World Scientific, Singapore, 1991).

\ref
[6]
Y.~Jack Ng and H.~van Dam, ``A small but nonzero cosmological constant'',
(Int. J. Mod. Phys D., to appear)
\eprint{hep-th/9911102}

\ref
[7]           
Djamel Dou,			
 ``Causal Sets, a Possible Interpretation for the Black Hole
 Entropy, and Related Topics'', 
 Ph.~D. thesis (SISSA, Trieste, 1999)

\ref
[8]		
Rafael D.~Sorkin,
``Indications of causal set cosmology'',
 {\it Int. J. Theor. Ph.} {\bf 39}(7): 1731-1736 (2000)
 (an issue devoted to the proceedings of the Peyresq IV conference,
  held June-July 1999, Peyresq France) 
\eprint{gr-qc/0003043}

\ref
[9] 
A.R.~Daughton,			
 {\it The Recovery of Locality for Causal Sets and Related Topics},
  Ph.D. dissertation (Syracuse University, 1993)   

\ref
[10] 
Rafael D.~Sorkin,
{\it A library of Lisp functions for posets and other purposes}, 
  \linebreak	
  \hbox{http://www.physics.syr.edu/$\widetilde{\phantom{n}}$sorkin}
  (Version 1.4, December 1998, Version 2.0 to be released soon)

\ref
[11] 
David P.~Rideout and Rafael D.~Sorkin,
``A Classical Sequential Growth Dynamics for Causal Sets'',
 {\it Phys. Rev. D} {\bf 61}, 024002 (2000)
 \eprint{gr-qc/9904062}

\ref
[12] E.B.~Dynkin, {\it Markov Processes} (Academic Press, 1965)

\ref
[13]
Xavier Martin, Denjoe O'Connor, David Rideout and Rafael D.~Sorkin,
``On the ``renormalization'' transformations induced by
  cycles of expansion and contraction in causal set cosmology'',
 {\it Phys. Rev. D} {\bf 63}, 084026 (2001)
\eprint{gr-qc/0009063} 

\ref
[14]  
John Stachel, ``Einstein's Search for General Covariance, 1912--1915'',
 in {\it Einstein and the History of General Relativity},
 edited by D.~Howard and J.~Stachel (Birkh{\"a}user 1989)
 %

\ref
[15]	
Edward Witten, 
``Quantum Gravity in De Sitter Space'',  
\eprint{hep-th/0106109}

\ref
[16]  
{H.~Fay Dowker}, {Raquel S.~Garc{\'\i}a}, {Joe Henson} and {Rafael D.~Sorkin}
(in preparation)

\ref
[17]
  R.D.~Sorkin,
``Quantum Measure Theory and its Interpretation'', 
  in
   {\it Quantum Classical Correspondence:  Proceedings of the $4^{\rm th}$ 
    Drexel Symposium on Quantum Nonintegrability},
     held Philadelphia, September 8-11, 1994,
    edited by D.H.~Feng and B-L~Hu, 
    pages 229--251
    (International Press, Cambridge Mass. 1997)
    \eprint{gr-qc/9507057}
  %

\end


(prog1    'now-outlining
  (Outline 
      "
     "......"
      "
   "\\\\message" 
   "\\\\section"
   "\\\\appendi"
   "\\\\Referen"	
   "\\\\Abstrac" 	
      "
   "\\\\subsectio"
   ))